\documentclass{article}
\usepackage{amsmath}
\usepackage{graphicx}
\usepackage{algorithm}
\usepackage{algorithmic}

\begin{document}

\title{DigiLock: User-controlled and Server-aware Digital Locker System}

%
%

\author{Atrayee Deb, Saloni Dalal, Manik Lal Das\vspace{2 mm}\\
DA-IICT\\
Gandhinagar, India\\
Email: \{atrayee\_deb, saloni\_dalal,
maniklal\_das\}@daiict.ac.in}

\maketitle

\begin{abstract}
The growing popularity of digital systems have paved the way for
digital locker that ensures security and safety of the digital
documents in store. While facilitating this system to user and
availing its services offered by service provider, non-repudiation
of service offered and service consumed is an important security
requirement in the digital locker system. In this paper, we
present a digital locker system that addresses the aspect of
\textit{confidentiality, integrity, and non-repudiation} along
with other security properties. The proposed protocol ensures the
confirmed participation of the user as well as the service
provider while accessing the digital locker. The protocol is
analyzed against potential threats in the context of safety and
security of the digital locker system.\vspace{1 mm}
\end{abstract}


\section{Introduction}
It has been witnessed that the exponential growth of mobile
devices along with the massive power of Internet has eased up the
ways of people living. Digital initiatives for facilitating daily
needs, whether it is for paying bills, banking, e-learning or
e-governance, are encouraging more and more users to come online
for saving time as well as money \cite{digi}. One of the efficient
services of this digital move is to have a digital store in cloud,
so that the user can access the service from anywhere and anytime.
Digital assets including personal and professional documents are
being stored on drive from where it can be accessed in minutes
irrespective of time and location. These storage facilities such
as Google drive \cite{goo}, Dropbox \cite{drop}, OneDrive
\cite{one} give us the facility to not only store the documents on
the cloud, but also to give different levels of access to
different people. This brings the notion of Digital Locker
\cite{mot}, \cite{bou} in which storing, accessing and ensuring
the authenticity and accountability of digital assets
have become much more efficient and secure.\vspace{1 mm}\\
Digital locker is a digital safe with the objective of keeping
important documents safe and secure under a mutually trusted
arrangement \cite{can}, \cite{elm}, \cite{kri}. The storage and
transfer of the documents take place mostly on an online platform,
which facilitates anytime-anywhere access to documents. One of the
important objectives of the digital locker is to ensure
non-repudiation of the parties involved as well as the
confidentiality and integrity of the documents \cite{elm}. Imagine
a scenario wherein documents stored in the locker are manipulated
by the user, and then the user can easily blame it on the service
provider denying its involvement in the act. Alternatively, the
service provider can also alter the documents and put the blame on
the user for being involved in the act of misdeed. Ideally, the
locker should be opened when both the user as well as the service
provider participate in the process, which usually happens in
operating of physical locker system \cite{kri} used in real-world
applications. Therefore, for any transaction with the locker both
the service provider and service consumer are equally responsible
and answerable for any misconduct intentionally or accidentally. A
typical scenario of conventional locker access system is depicted
in Figure 1.

\begin{figure}[ht!]
\centering
\includegraphics[width=9 cm, height=7 cm]{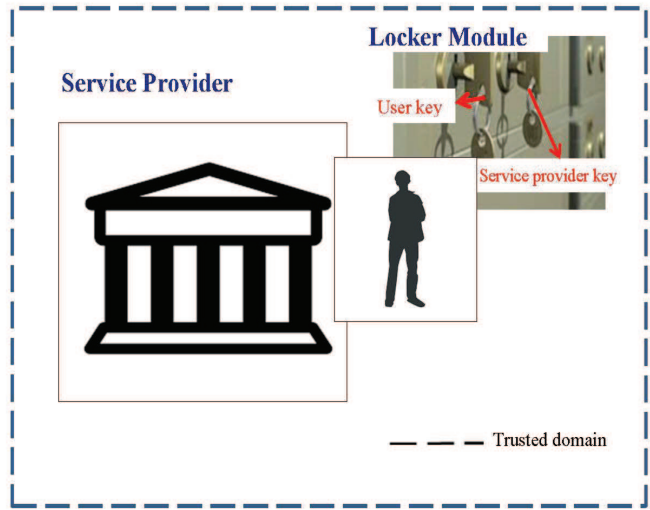}
\caption{Conventional Locker System}
\end{figure}
\noindent \textbf{Our contributions:} In this paper, we present a
protocol for securing digital locker system that ensures the
locker access is given to the user only on the participation of
both the user key and the service provider key. All necessary
validations of user and service provider participation takes place
inside the locker system, which is managed securely by the service
provider's storage system. A trusted module acts as the safe
locker, performs all operations after checking credentials of
user's request and service provider's consent. The proposed
protocol is analyzed against potential threats in the context of
digital locker system and shown secure against repudiation, replay
and impersonation attacks.\vspace{1 mm}\\
\textbf{The organization of the paper:} The remaining of the paper
is organized as follows. Section 2 presents the protocol for
securing digital locker system. Section 3 analyzes the proposed
protocol. We conclude the paper in Section 4.

\section{Digital Locker System}

\subsection{System Model and Design Goal}
The system consists of three entities -- User, Service Provider
and the Locker Module. The assumptions and functionalities of
these entities are listed below:\vspace{1 mm}\\
\textbf{User:} User is required to register at the service
provider for availing the services of the locker system. For this,
the user is prompted to submit a secret key $K_i$ and a secret
question $m$. The user chooses $m$ to be something that the user
can remember at a later time for logging into the system. The
secret $m$ is encrypted and the secret key $K_i$ is hashed and
stored at the locker system. User secret key $K_i$ is not known to
the
locker system.\vspace{1 mm}\\
\textbf{Service Provider:} The service provider uses the secret
key $R$ to connect to the locker. Without service provider's
involvement the user cannot get access to the locker. All the
functions performed and all the information stored within the
locker system are assumed to be unavailable to the service
provider's server. The master secret key $R$ of the service
provider is not known to the user.\vspace{1 mm}\\
\textbf{Locker Module:} This functionality ensures the
participation of the parties, both user and the service provider.
The system has $h(K_i)$ and $h(R)$ stored in it for verification
of the secret keys of user and service provider, respectively. It
also stores the user secret $m$ appended with user secret key
$K_i$ encrypted as $E_{L}(m\|K_i)$, where $L$ is a computed
symmetric key.

\begin{figure}[ht!]
\centering
\includegraphics[width=8 cm, height=6 cm]{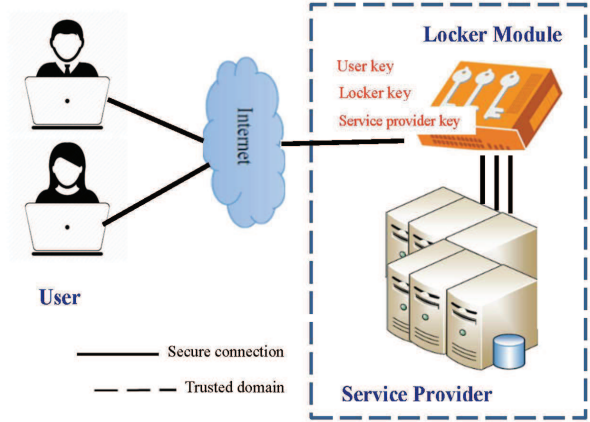}
\caption{The System Model}
\end{figure}
\noindent \textbf{Design Goal.} There are two main design
objectives of the proposed protocol. Firstly, there should be
participation of both the entities for accessing the locker just
like that in a physical locker system. For example, a physical
locker in a bank requires user to insert the user's key as well as
the bank's key. Later, in case of any theft or misconduct, the
bank cannot back out denying its participation nor can the user
put false allegations on the bank. Thus, ensuring non-repudiation
of both parties in the digital locker system is an integral
requirement. Secondly, the user can verify the computation results
and can send an acknowledgement message to the locker module. If
the acknowledgement is correct, then only the locker opens and the
user is allowed access. In other words, the locker module ensures
the usage of user's key and service provider's key before opening
the digital lock for accessing user's digital assets.\vspace{1 mm}\\
The symbols and notations used in the protocol are described in
Table \ref{tab:not}.

\begin{table}[!htbp]
\centering
\begin{tabular}{p{2 cm}||p{9 cm}}
\hline
& \\
Notation & Description\\\hline \hline
& \\
$R$ & the master secret key of the bank which will be needed to
ensure the participation of the bank. \\
$K_i$ & the key of $i^{th}$ user. \\
$N_a$ & the random number generated by the user for each session. \\
$m$ & the secret question chosen by the user during the time of
registration. \\
$N_r$ & the random number generated by the locker facility for
each session. \\
$K_s$ & the shared symmetric key between user and the locker
system. \\
$L$ & the symmetric key with which locker encrypts the correct
version of user key and the secret message. \\
$E_{x}(message)$  & the message is encrypted with $x$. \\
$D_{x}(message)$ & the message in decrypted with $x$. \\
\hline
\end{tabular}
\vspace{3 pt} \caption{Symbols and Notations used in the Protocol}
\label{tab:not}
\end{table}
We note that $m$ is a secret question that must be remembered by
the user without storing it. Upon successful validation of the
user into the locker system, the user can see this $m$ on the
screen.

\subsection{The Proposed Protocol}
The protocol consists of a one-time Setup phase, and Locker Access
phase as when user wants to access the locker system.

\subsubsection{Setup phase}
The user requests to connect to the service provider server. The
connection to the service provider server is done through a secure
TLS connection over HTTPS \cite{tls}, \cite{mld}. Once secure
session is established between user and service provider, the user
requests access to the locker facility of the service provider.
All the messages between the user and the locker system are sent
through the service provider server (e.g. like bank acts the
service provider between the user and the physical locker in
conventional locker system). Each user, say $U_i$, selects a
different secret key $K_i$ for accessing the locker system. The
service provider has a master key $R$ known to service provider
server only. It is assumed that the secret keys of the users and
the master key of the service provider remain uncompromised. The
locker system stores the hash of the $<$ user\_id, user key$>$,
that is $h(U_i\|K_i)$ as well as the hash of the server key
$h(R)$. Each user submits a secret question $m$ to the server at
the time of registration. The secret $m$ is stored in the locker
system corresponding to each user. The secret $m$ is a word or a
phrase that is easy for the user to remember and verify. The
locker system computes a symmetric key $L$ as $h(U_i\|K_i)$
$\oplus$ $h(R)$. The user secret message $m$, and $K_i$ are stored
in the locker system encrypted with this symmetric key $L$ as
$E_L(m\|K_i\|U_i)$. All other message exchanges are encrypted with
the help of the shared secret key $K_s$ as computed in the
\textit{Verification and Access of Locker} phase explained below.

\subsubsection{Locker Access phase} This phase is comprised of
User-Locker Authentication and Locker Access sub-phases.\vspace{1
mm}\\
\textit{User and Locker Authentication:} User $U_i$ generates a
random number $N_a$. The value of a pseudo-random function
$PRF_{h(U_i\|K_i)}(N_{a})$ is calculated with $h(U_i\|K_i)$ as the
key. The user sends $PRF_{h(U_i\|K_i)}(N_{a})$ and $N_a$ to the
locker system. When $PRF_{h(U_i\|K_i)}(N_{a})$ and $N_a$ are sent
through the service provider server, the locker system verifies
$h(U_i\|K_i)$ by computing $PRF_{h(U_i\|K_i)}(N_{a})$ with its
pre-stored value of $h(U_i\|K_i)$. If the computed value does not
match with the received one, then the user is notified with an
error message stating that the key submitted was incorrect. If the
values match, then the process is carried forward. If the user is
verified, the locker system requests for the service provider
master key $R$. On getting the service provider master key, the
locker system calculates $h(R)$ and matches it with the pre-stored
value. In case of mismatch of this verification, the
authentication of service provider cannot be established, and as a
result, the user's request gets denied. The instruction-flow of
the locker access system is shown in Figure 3.

\begin{figure}[htbp]
\centering
\includegraphics[width=13 cm, height=6 cm]{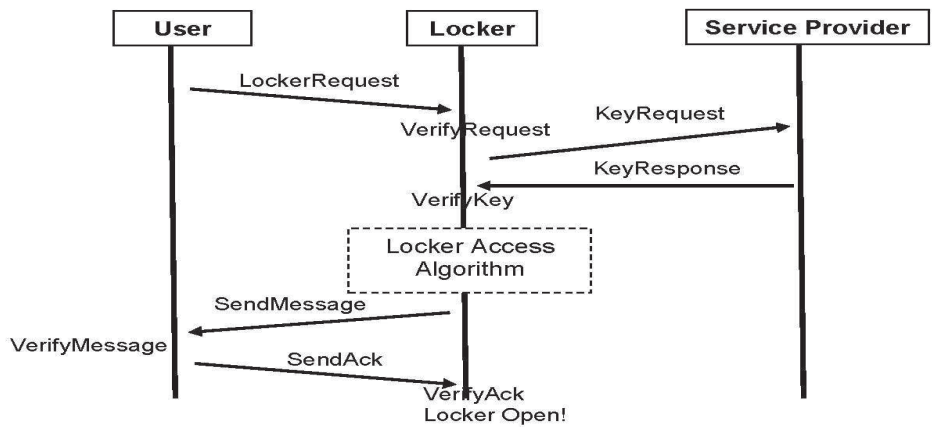}
\caption{The work-flow of the Locker System Access}
\end{figure}

\textit{Locker Access:} Following operations are performed to
ensure that both user and the service provider participate in
accessing the locker system.

\newcounter{r1}
\begin{list}{Step-{\arabic{r1}.}}
{\usecounter{r1}} \item The user symmetric key $L$ is calculated
as $L$ = $h(U_i\|K_i)$ $\oplus$ $h(R)$, which is used to decrypt
the pre-stored encrypted message $E_{L}(m\|K_i\|U_i)$. On
decrypting the message, $K_i$ is extracted from
$(m\|K_i\|U_i)$.\vspace{2 mm}

\item Locker system then computes the shared secret key $K_s$ as
$K_{s}$ = $h(U_i\|K_i\|N_a$) using the user secret key $K_i$ and
the random number $N_a$ sent by the user.\vspace{2 mm}

\item The Locker generates a random number $N_r$.\vspace{2 mm}

\item The secret message $m$ and $N_r$ have to protected while
transmitting them to the user. The locker system encrypts the user
secret message $m$ and $N_r$ with $K_s$ and sends it to the user.
\end{list}
Once the user receives the message from the locker, it also
computes the symmetric key $K_s$. It uses the freshly computed
$K_s$ to decrypt the received message and gets the secret message
$m$. This secret message is verified by the user. If the secret
message $m$ is displayed correctly it ensures correctness of the
protocol as follows: (i) the server master key $R$ was submitted
correctly by the service provider server and hence also ensuring
its participation, and (ii) since the $K_s$ is computed using the
random number $N_a$ given by the user, it ensures the freshness
property. After confirming the correctness of the secret message
$m$, the user sends an acknowledgement to the locker system. It
computes a digest $h(N_a\|N_r)$ with $N_r$ and $N_a$ as inputs and
transmits it to the locker system. The locker already has $N_a$
and $N_r$ and thus it verifies this digest by recomputing
$h(N_a\|N_r)$. This is done to ensure the user's agreement to
access the locker services. Once the verification is succeeded,
the locker services are made available to the user. If the secret
message $m$ is not decrypted properly, that is, if not confirmed
by the user then one of the above statements is invalid and hence
user does not send back the acknowledgement. If the locker does
not receive the confirmation within certain time period it ends
the current session without opening the locker to the user. After
the successful authentication of the user-locker system, the
Locker Access works with Algorithm 1.

\begin{algorithm}[H]
\caption{Locker Access}
\begin{algorithmic}[1]
\STATE Locker computes user's key $L$ as $L$ = $h(U_i\|K_i)$
$\oplus$ $h(R)$.\vspace{1 mm}

\STATE Locker decrypts $E_{L}(m\|K_i\|U_i)$ using the key $L$ and
obtains $m$ and $K_i$ for the user.\vspace{1 mm}

\STATE Locker computes the shared secret key $K_s$ =
$h(U_i\|K_i\|N_a$), where $N_a$ is provided by the user.\vspace{1
mm}

\STATE Locker generates a random number $N_r$.\vspace{1 mm}

\STATE Locker encrypts $E_{K_s}(m\|N_r)$ and sends it to the user.
\vspace{1 mm}

\STATE User decrypts $E_{K_s}(m\|N_r)$ and gets the secret message
$m$.\vspace{1 mm}

\STATE User sends an acknowledgement $h(N_a\|N_r)$ to the
Locker.\vspace{1 mm}

\STATE Upon receiving user's acknowledgement the locker verifies
it and allows the user to access the Locker space allocated for
the user.

\end{algorithmic}
\end{algorithm}

The DigiLock protocol is shown in Figure 4.

\begin{figure}[ht!]
\centering
\includegraphics[width=10cm, height=11cm]{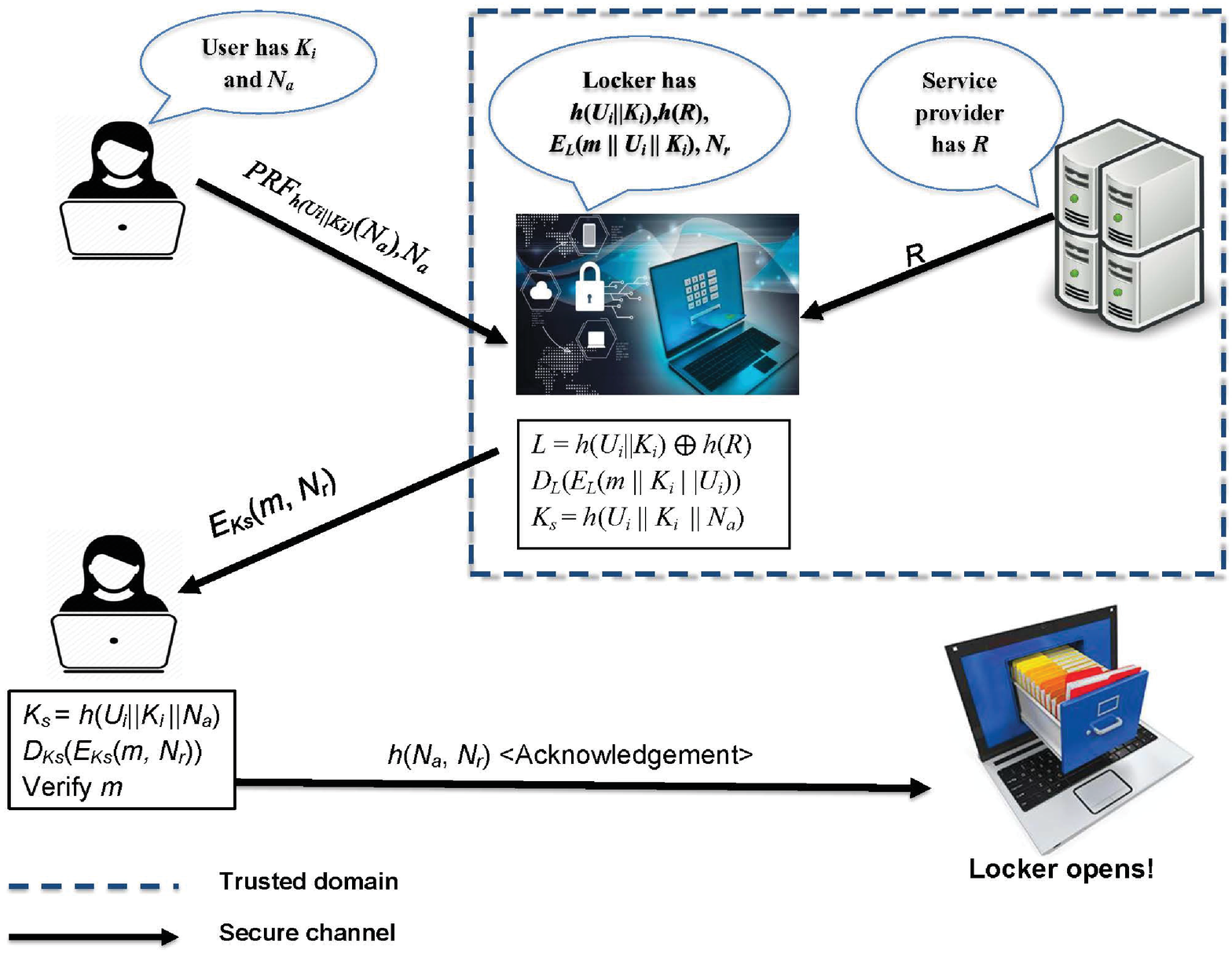}
\caption{DigiLock: The proposed protocol}
\end{figure}

\section{Security Analysis}
We note that the proposed protocol for locker system is run under
the protection of TLS protocol. In other words, the message
exchange between the participating entities in the proposed locker
system is done through a secure channel. On top of the TLS
protection, we show that the proposed locker system is secure
against repudiation, replay and impersonation attacks.\vspace{1 mm}\\
\textbf{Security against repudiation.} Non-repudiation being a
major issue in digital locker system, is addressed in our protocol
as it ensures the participation of both the entities the user and
the service provider, so that at a later time none of them can
back out from their participation in the accessing of the locker.
The encryption key $L$ is computed using both $K_i$ (key of the
user) and $R$ (key of the service provider). Assume that the user
is sending incorrect $K_i'$. Then the computation of encryption
key will be as follows.

$L'$= $h(U_i\|K'_i)$ $\oplus$ $h(R)$

$D_L'$($E_L$($m\|K_i\|U_i$))\vspace{1 mm}\\
Since the encryption is done with the key that involved the true
$K_i$, the message will not be decrypted properly. As a result,
the message $m$ will not be displayed on the screen and thereby,
locker will not open. Same is the case if the service provider
enters an incorrect $R$. If either of them do not participate or
provide a faulty input, then the message $m$ is not decrypted
properly and the locker will not open. Therefore, for availing the
locker services, the message $m$ has to be displayed correctly for
which the participation of both user and the service provider is
essential.\vspace{1 mm}\\
\textbf{Security against replay attack.} If an attacker tries to
replay messages intercepted from previous run of the protocol to
make locker system believe that its a genuine user, then the
attacker will not succeed because the locker system always
calculates $K_s$ with the fresh nonce $N_a$ generated by the user.\\
Suppose the attacker $\mathcal{A}$ intercepts a message from an
older session and uses it in the current session to manipulate the
user $U$ and the locker server $L$, i.e., replays an older message
to manipulate the user or locker server into believing that they
are communicating with the each other.

$U \rightarrow \mathcal{A}: PRF_{h(U_i\|K_i)}(N_a), N_a$

$\mathcal{A} \rightarrow L: PRF_{h(U_i\|K_i)}(N_a'),
N_a'$\vspace{1 mm}\\
As the $N_a$ inside the function matches with the external one,
locker thinks that it has got message from right user and
authenticates the sender. It then does the computations as
follows.

$K_s' = h(U_i\|K_i\| N_a')$

$L \rightarrow \mathcal{A} : \{m, N_r\}K_s'$

$\mathcal{A} \rightarrow U : \{m, N_r\}K_s'$

$U$ computes the key $K_s = h(U_i\|K_i\| N_a)$\vspace{1 mm}\\
The secret message $m$ does not get decrypted properly as $K_s'
\ne K_s$ and hence the decrypted message $m' \ne m$. Therefore,
the user does not send the acknowledgement message to the locker,
thereby, mitigating the replay attack.\vspace{1 mm}\\
\textbf{Security against impersonation attack.} The user key $K_i$
is kept confidential and uncompromised, which prevents any
adversary to impersonate the user. Suppose that the server $S$
tries to impersonate the user, i.e., tries to imitate the
behaviour of the user without hinting the locker system about it
by computing the following:

$U  \rightarrow S: PRF_{h(U_i\|K_i)}(N_a), N_a$

$S \rightarrow L: PRF_{h(U_i\|K_i)}(N_a), N_a$\vspace{1 mm}\\
Now since the $N_a$ inside the function matches with the external
one, locker thinks that it has got message from right user and
authenticates the sender. It then computes

$L \rightarrow S: \{m, N_r\}_{K_s}$\vspace{1 mm}\\
Now, $S$ that is for example, the service provider cannot decrypt
the message as it cannot compute $K_s$ because it does not have
$K_i$. No acknowledgement message will be sent to the locker. As a
result, locker impersonation by the server is prevented in our
protocol. Furthermore, suppose that the service provider tries to
imitate the locker system. As we have stated above, we assume that
all the information stored in the locker is concealed from the
service provider server. Thus, the service provider will not have
access to the encrypted message. Therefore, it cannot send the
message back to the user for further verification.

\section{Conclusion} Digital lockers are becoming popular in
modern digital space because of its ease of use from anywhere and
at anytime. We have proposed a locker system that ensures the
participation of both the user and the server so as to ensure
non-repudiation of participatory role of availing as well as
facilitating this service. In addition, the proposed protocol
ensures the data confidentiality and integrity during the message
exchange of the protocol. The proposed protocol mitigates
impersonation, replay attacks, and ensures that any illegitimate
party is not able to breach the security of the locker system.

\end{document}